\documentclass[prb]{revtex4}
\pdfoutput=1
\usepackage{graphicx}
\usepackage{epsfig}
\usepackage{ae}
\usepackage{aecompl}
\usepackage{amsmath}

\begin{document}

\title{Ionic Effects on the Electric Field needed to Orient Dielectric 
Lamellae}

\author{G. Garb{\`e}s Putzel$^1$, David Andelman$^2$, Yoav Tsori$^3$ and
  M. Schick$^1$ }
\affiliation{$^1$Department of Physics, University of Washington,  Box
  351560, Seattle, WA 98195-1560 \\
$^2$Raymond and Beverly Sackler School of Physics and Astronomy,
Tel Aviv University, Ramat Aviv 69978, Tel Aviv, Israel\\
$^3$Department of Chemical Engineering, Ben Gurion University of the
  Negev, P.O. Box
  653, Beer Sheva 84105, Israel}

\begin{abstract}
We consider the effect of mobile ions on the applied potential needed to 
reorient a lamellar system of two different materials placed between two 
planar electrodes. The reorientation occurs from a configuration parallel 
to the electrodes favored by surface interactions to an orientation 
perpendicular to the electrodes favored by the electric field. The system 
consists of alternating A and B layers with different dielectric 
constants. The mobile ions are assumed to be insoluble in the B layers and 
hence confined to the A layers. We find that the ions reduce the needed 
voltage most strongly when they are constrained such that each A lamella 
is electrically neutral. In this case, a macroscopic separation of charge 
and its concomitant lowering of free energy, is attained only in the 
perpendicular orientation. When the ions are free to move between 
different A layers, such that charge neutrality is only required globally, 
their effect is smaller and depends upon the preferred surface interaction 
of the two materials. Under some conditions, the addition of ions can 
actually stabilize the parallel configuration. Our predictions are 
relevant to recent experiments conducted on lamellar phases of diblock 
copolymer films with ionic selective impurities.
\end{abstract}
\maketitle

\newpage

%%%%%%%%%%%%%%%%%%%%%%%%%%%%%%%%%%%%
\section {INTRODUCTION}
%%%%%%%%%%%%%%%%%%%%%%%%%%%%%%%%%%%%

Block copolymers are polymeric materials with specific chain 
architecture in which several blocks of different chemical nature are 
linked covalently. The simplest design is that of a di-block 
in which two polymeric blocks, A and B, are connected. Even in this design 
the system manifests a rich phase structure of self-assembled periodic 
arrays of spheres, cylinders, or planes, which has many applications such 
as templates for nanoscopic devices 
\cite{park97,walheim99,morkved96,talbrechtsci00,lopes01,park03}. One 
difficulty which must be overcome before block copolymer arrays can be 
exploited is the fact that, when produced as a thin film, the array tends 
to align with the substrate due to preferential interactions between it 
and one of the polymer blocks. In many applications, however, the desired 
orientation is one in which the ordered structure is aligned perpendicular 
to a substrate. For example, if a cylindrical array is aligned 
perpendicularly, then the cylinder cores can be etched away and replaced 
by a metal. Further removal of the material surrounding the cores produces 
an array of wires useful for sensors \cite{park97}. One method to bring 
about this orientation is to subject the film to an electric field 
\cite{amundson93,amundson94,morkved96,pereira99,boker02,tsori02,ashok01}, 
a technique which takes advantage of the difference in dielectric 
constants between the two polymer blocks. Because the electric field 
needed to produce the desired orientation can be rather high, on the order 
of $10$\,V/$\mu$m, it can happen that the material undergoes dielectric 
breakdown before alignment is achieved. As a way of lowering the electric 
field needed for alignment, Tsori et al.~\cite{tsori03} suggested that one 
could introduce free ions into the system. Presumably if the ions were 
preferentially soluble in one of the blocks, the induced dipole moment of 
that block would become much larger in the perpendicular orientation. The 
increase in polarization would bring about the desired decrease in the 
electric field needed for alignment.

This procedure has been utilized experimentally and the desired increase 
in orientation obtained for a given applied field \cite{xu04,wang06}. 
In spite of this success, it is not completely clear how the addition of 
ions brings about the desired result. It has been argued that instead of 
increasing the induced dipole moment, large amounts of added ions 
introduced in one of the two blocks simply change the respective 
dielectric constant\cite{wang06}. A different argument \cite{wang08} is 
that the addition of large amounts of ions changes the strength of the interaction between the 
different monomers even in the absence of an external field. In support of 
this latter suggestion, introduction of ions has been shown to affect the 
phase behavior of block copolymers, behavior which scales with this 
interaction strength \cite{ruzette01,epps02}.

To elucidate the origin of free--ion effects in block-copolymer systems, 
we have 
considered the effect of free ions on a system which is simpler, 
but which shares its mesoscopic periodic morphology. Our model 
system is a rigid stack of alternating A/B lamellae. The entire stack is 
confined between two external flat electrodes. The A and B lamellae are 
characterized by two dielectric constants, $\kappa_A$ and $\kappa_B$, 
respectively. Free ions are then introduced into the A layers. It is 
assumed, arbitrarily, that the positive charges are immobile and 
distributed uniformly throughout the A layers, creating a uniform charged 
background, while the negative counterions are mobile throughout the A 
regions. An interchange of the role of positive and negative charges will 
have no effect on the results of our calculation. The fraction of ions 
dissolved in the two kinds of layers is, in general, different, 
and we consider the 
case of greatest contrast in which there are no charges in the B lamellae. 
We calculate, within mean-field theory, the system free-energy in the 
presence of a voltage $V_0$ applied across the plates. The free energy 
depends, of course, upon the relative orientation of lamellae and electric 
field.

Our results for the critical reorienting electric field depend strongly 
upon the location of the counterions. If they are confined to the A lamellae 
such that each is electrically neutral, then separation of charge 
occurs only over distances of the order of the thickness of the A lamellae 
when the lamellae are parallel to the substrate. In contrast, the charges 
can be separated over lengths of the order of the entire film thickness 
when lamellae are oriented perpendicularly. Hence, the latter 
configuration is favored, and the desired orientation can be brought about 
with an electric field significantly weaker than that needed without ions. 
However, if the counterions are free to move between different A lamellae, 
then the separation of charge occurs over distances of the order of the 
sample thickness in both orientations, so that the system behavior is not 
easily predicted.  In this case, our results show that there is a smaller reduction of the field needed to 
reorient the sample, and that the magnitude of the reduction 
depends upon whether the surface interactions prefer the A component,
whose lamellae contain ions, or the B 
component. The effect is larger if the surface interactions 
prefer the B component.  
If they prefer the A component, we find that the introduction of ions can 
even stabilize the parallel orientation.

%%%%%%%%%%%%%%%%%%%%%%%%%%%%%%%%%%%%
\section{The Model}
%%%%%%%%%%%%%%%%%%%%%%%%%%%%%%%%%%%%

We consider a system, of volume $\Omega$, consisting of alternating layers 
of materials denoted A and B with dielectric constants $\kappa_A$ and 
$\kappa_B$, respectively. All layers but the two closest to the two 
surfaces (chosen arbitrarily to be B layers) are of equal thickness 
$\lambda/2$ so that the structural periodicity is $\lambda$. The two 
layers adjacent to the plate electrodes are each of thickness $\lambda/4$, 
as they would be in a system of { unstrained} lamellar block copolymer. 
The material 
fills the space between two plates, which are parallel to the ($x,y$) 
plane, one at $z=0$ at which $V(x,y,0)=0$, the other at $z=d$ at which 
$V(x,y,d)=V_0.$ Each plate is of area ${\cal S}_{\rm plate}$. Just as in 
the experiments, there is a thin insulating layer between the plates and 
the material, keeping the total amount of ions in between the two plates 
fixed. We assume the system contains $N_+$ immobile, monovalent, cations 
distributed with uniform density $\rho_+$ in the A lamellae, and zero 
density within the B lamellae. The counterions are also monovalent and 
assumed to be mobile with a non-zero number density, $\rho_-({\bf r})$, 
only within the A lamellae; there are no counterions in the B lamellae, 
and the entire lamellar system is electrically neutral.

The Maxwell equations which govern this electrostatic system with linear
dielectric materials are, within the SI system,
\begin{eqnarray}
\label{max1}
\nabla\times{\bf E}({\bf r})&=&0 \qquad {\rm and}\\
\label{max2}
\nabla\cdot (\kappa({\bf r})\epsilon_0{\bf E}({\bf r}))&=&
e(\rho_+({\bf r})-\rho_-({\bf r})),
\end{eqnarray}
where $e>0$ is the unit of charge. The first equation is satisfied
identically by introducing the electrostatic potential $V({\bf r})$
such that ${\bf E}({\bf r})=-\nabla V({\bf r}).$
With this the remaining Maxwell equation takes the form

\begin{equation}
 \label{modpoisson}
 \nabla \cdot(\kappa({\bf r})\epsilon_0\nabla V({\bf r}))=-e(\rho_+({\bf
r})-\rho_-({\bf r})).
\end{equation}
In our model system, the number density of positive charges within the A
lamellae is constant, $\rho_+({\bf r})=\rho_+$, and the dielectric
constant $\kappa({\bf r})=\kappa_A$, a constant, so that
Eq.~(\ref{modpoisson}) becomes the Poisson equation
\begin{equation}
\label{poisson}
\nabla^2V({\bf r})=-\frac{e}{\kappa_A\epsilon_0}(\rho_+-\rho_-({\bf
r}))\qquad {\bf r}\ {\rm in\ lamellae\  A}.
\end{equation}
Within the B lamellae, the number density of all charges, cations and anions,
is taken to vanish, and $\kappa({\bf r})=\kappa_B$ a constant, so
that within these regions Eq.~(\ref{modpoisson}) becomes the
Laplace equation
\begin{equation}
\label{laplace}
\nabla^2 V({\bf r})=0\qquad {\bf r}\ {\rm in\ lamellae\ B}.
\end{equation}
At an A/B interface at which the dielectric properties of the material 
changes abruptly, the potential $V({\bf r})$ is continuous, the parallel 
component of ${\bf E}$ is continuous as follows from Eq.~(\ref{max1}), and 
the normal component of the displacement field, ${\bf 
D}\equiv\kappa\epsilon_0{\bf E},$ is continuous as follows from 
Eq.~(\ref{max2}). The potential is now completely specified as a 
functional of the unknown charge density,
\begin{equation}
\label{pot}
V({\bf r})=V[\rho_-({\bf r})].
\end{equation}

Within mean-field theory, the relation between the ensemble-averaged charge
density and potential satisfies the Boltzmann distribution:
\begin{equation}
\label{mfdensity}
\rho_-({\bf r})=\rho_+\Omega\frac{
\exp[\beta e V({\bf r})]}{\int_Ad{\bf r}\exp[\beta eV({\bf r})]}\qquad {\bf r}\
    {\rm in\ A},
\end{equation}
with $\beta\equiv 1/k_BT$, $k_B$ is the Boltzmann constant, $T$ is the
absolute temperature, and the integral is taken over all the A lamellae.

Equations (\ref{poisson}), (\ref{laplace}), and (\ref{mfdensity}) 
constitute the self-consistent, Poisson-Boltzmann, equations which must be 
solved subject to the boundary conditions given above. Once the 
solutions for the charge density and potential are determined, the free 
energy of the system can be obtained. It is convenient to display 
explicitly the contribution to it from the surface fields and to write the 
total free energy as
\begin{equation}
\label{fsurf}
F_{\rm tot}(T,\Omega,V_0,N_+,{\cal S}_A,{\cal S}_B)=
F_{\rm el} + \gamma_A{\cal S}_A+\gamma_B{\cal
  S}_B,
\end{equation}
where ${\cal S}_A$ is the total area of the two plates
in contact with the A region and similarly for ${\cal S}_B$; $\gamma_A$
and $\gamma_B$ are the interfacial free energies per unit area between the
plates and A and B regions, respectively;
${\cal S}_A+{\cal S}_B=2{\cal S}_{\rm plate}.$

For convenience, we remind the reader in Appendix I of the result for the 
electrostatic energy of the system when the plates are held at a fixed 
potential difference. With the energy in hand, the mean-field expression 
for the electrostatic part of the free energy can be derived in various 
ways \cite{edwards65,borukhov00,fredrickson06} with the result
\begin{eqnarray}
\label{fmft}
F_{\rm el}(T,\Omega,V_0,N_+,{\cal S}_{\rm plate})
&=&-\frac{1}{2}\epsilon_0\int\kappa({\bf r})(\nabla V({\bf
r}))^2d{\bf r}+e\int_A V({\bf r})(\rho_+-\rho_-({\bf r}))d{\bf r} \nonumber \\
                    &+&k_BT\int_A
\rho_-({\bf r})\ln(\rho_-({\bf r})/\rho_+)\,d{\bf r}.
\end{eqnarray}

The free energy is proportional to the area of each plate, so it is 
convenient to consider the free energy per unit area. We choose to measure 
the free energy per area in units of $2\epsilon_0\kappa_A/\beta^2\lambda 
e^2$. Hence we introduce dimensionless free energies per unit area,
\begin{eqnarray}
 f_{\rm tot}(T,\rho_+,V_0,{\cal S}_A/{\cal S}_{plate})&\equiv& 
\frac{\beta^2\lambda
e^2}{2\epsilon_0\kappa_A}\frac{ F_{\rm tot}}{{\cal
S}_{\rm plate}},\\
\label{scaledfel}
f_{\rm el}(T,\rho_+,V_0)&\equiv&\frac{\beta^2\lambda
e^2}{2\epsilon_0\kappa_A}\frac {F_{\rm el}}{{\cal
S}_{\rm plate}},
\end{eqnarray}
with $F_{\rm el}$ given by Eq.~(\ref{fmft}),
so that within mean-field theory, Eq.~(\ref{fsurf}) becomes
\begin{equation}
 \label{dimless}
f_{\rm tot}= f_{\rm el}+ \left(\frac{\beta^2\lambda
e^2}{2\epsilon_0\kappa_A}\right) \left(\frac{{\cal S}_A}{{\cal
S}_{\rm plate}}\gamma_A+\frac{{\cal S}_B}{{\cal S}_{\rm plate}}\gamma_B\right).
\end{equation}

To determine the orientation of the lamellae in the presence of the 
external field, we determine the free energy for the two orientations: 
lamellae that are perpendicular to the substrate and those that are 
parallel to it. In the latter case, we approximate the situation expected 
if the system were an unstrained A/B diblock copolymer film. As the 
substrate prefers one of the two blocks, the lamellae next to the two 
plates consist of the preferred block (assumed here to be the B block) and 
will have a thickness of $\lambda/4$, while all other A and B lamellae are 
of thickness $\lambda/2$.

We consider two different constraints on the location of the mobile 
counterions. In the first, we assume that the counterions are not free to 
move between A lamellae. Their distribution is such that {\em each} A 
lamella is electrically neutral. We refer to this as {\em local 
neutrality}. The other possibility we consider is that the counterions, 
while found only in the A lamellae as before, are distributed among them 
subject only to the weaker constraint that the system is {\em overall} 
neutral.  We refer to this as {\em global neutrality}. The free energy of 
the system in which the lamellae are oriented parallel to the plates is 
affected significantly by the difference between these two constraints, 
while the system of perpendicular lamellae is not affected at all.

Before discussing the solution of the equations, we note that there are a
few dimensionless ratios which parametrize them. The simplest of these is
a dimensionless applied potential, $v$
\begin{equation}
v\equiv e\beta V_0.
\end{equation}
As typical potentials across the film are a few volts to a few dozens 
volts, this parameter is large, on the order of $10^2$ for experiments at 
room temperature. A dimensionless measure, $r$, of the ionic number 
density $\rho_+$ is readily defined
\begin{equation}
\label{rr}
r\equiv\left(\frac{\beta e^2\rho_+ d^2}{\kappa_A\epsilon_0}\right)^{1/2},
\end{equation}
and is easily seen to be the ratio of the system thickness
 $d$ to the Debye length, $\lambda_D$
\begin{equation}
        r=\frac{d}{\lambda_D}
\end{equation}
where
\begin{equation}
        \lambda_D=\left[\frac{\kappa_A\epsilon_0}{\beta
        e^2\rho_+}\right]^{1/2}.
\end{equation}
Thus, small values of $r$ imply weak screening of the
electric field by the counterions, while large values of $r$ imply strong
screening. The temperature-independent ratio
\begin{equation}
\label{r2v}
\frac{r^2}{v}=\frac{e\rho_+d^2}{\kappa_A\epsilon_0 V_0}\sim 
\frac{\rho_+}{V_0} ,
\end{equation}
depends only on the ratio $\rho_+/V_0$, which indicates that
the behavior in the limit of large externally
applied potential is the same as for small ionic charge density.

It is convenient to write the Poisson-Boltzmann equation in dimensionless 
form. To this end we measure all distances in units of $d$, the distance 
between the two bounding electrodes, ${\tilde{\bf r}}\equiv {\bf r}/d,$ 
and define the dimensionless potential $W({\tilde x},{\tilde y},{\tilde 
z})\equiv \beta e V(x,y,z)$ so that $W({\tilde x},{\tilde y},0)=0$, 
$W({\tilde x},{\tilde y},1)=v.$ Then the Poisson-Boltzmann equation takes 
the form
\begin{eqnarray}
\label{pb}
{\tilde\nabla}^2W({\tilde x},{\tilde y},{\tilde
  z})&=&-r^2\left(1-\theta{\rm e}^{W({\tilde z})}\right) \qquad 
\ {\rm in\  lamellae\ A,}\\
\theta&=&\frac{\rho_-(W{=}0)}{\rho_+},
\end{eqnarray}
and, of course, Laplace's equation
\begin{equation}
\label{laplace4}
{\tilde\nabla}^2W({\tilde x},{\tilde y},{\tilde z})=0\qquad  
{\rm in\  lamellae\ B}.
\end{equation}

We obtain estimates of the parameter $r$, and therefore the density of 
free ions from experiments on polystyrene-polymethylmethacrylate (PS-PMMA) 
diblock copolymers where lithium salts are infused in the MMA blocks. It is known that 
Li$^+$ ions are associated with the carbonyl groups of the MMA (dielectric constant
$\kappa_A=6.3$). 
The number density of the MMA monomers is about $7\times 10^{27}$m$^{-3}$. What is 
known less well is the number density of Li$^+$ ions solubilized in the 
MMA blocks. Wang et al. \cite{wang08} estimate that the largest fraction 
of Li$^+$ is about 0.27, which would correspond to a number density
$\rho_+\simeq 2\times 10^{27}$m$^{-3}.$ If all this charge were mobile, 
it would yield a nominal Debye length
$\lambda_D=0.067$nm at $T=298\,K$. A film thickness of 300nm as in Xu et al. \cite{xu04}, 
would imply a large value of $r\approx 4500.$ However, such a sub-atomic 
value for 
$\lambda_D$ would indicate that continuum theories, such as ours, would
be inapplicable.
In other experiments, the fraction of ions is clearly much smaller. For example, 
Kohn et al. \cite{kohn09}
quote a Debye length, $\lambda_D=2.3$ nm which, for a film of the same thickness, 
would yield $r\approx130.$ Further Tsori et al.  \cite{tsori03} estimate the  
the fraction of residual Li$^+$ ions to MMA  to be much smaller than 
in the work of Wang et al. \cite{wang08}, about 
3$\times 10^{-5}$. This means that $\rho_+\approx 2\times 
10^{23}$m$^{-3},$ corresponding to a Debye length of 6.7 nm. With the same film thickness 
of 300nm,  $r\approx 45.$ For films only a few lamellae in thickness, $r$ would be even smaller.
Thus the parameter $r$ is expected to vary between values greater than, but of 
order of unity to those as large as the order of $10^3$. For the smaller 
values in this interval, one expects the effect of the free ions to be but 
a perturbation on the ion-free results, while this is certainly not 
expected to be the case for the larger values. Thus, experiments span a 
range over which the effect of the ions is expected to vary from 
negligible to important.

%%%%%%%%%%%%%%%%%%%%%%%%%%%%%%%%
\section{Results}
%%%%%%%%%%%%%%%%%%%%%%%%%%%%%%%%%

%%%%%%%%%%%%%%%%%%%%%%%%%%%%%%%%%%
\subsection{The Parallel Orientation}
%%%%%%%%%%%%%%%%%%%%%%%%%%%%%%%%%%%%

We first consider the case in which the lamellae are parallel to the 
plates. The total amounts of A and B material are equal, and the B layers 
have no charges. Were there also no charges in the A layers, the electric 
fields would be
$E_A=2V_0/[(1+\kappa)d]$ and $E_B=\kappa E_A$ with $\kappa\equiv
\kappa_A/\kappa_B$. The potential would increase linearly across the
lamellae. The
magnitudes of the dimensionless electric fields, $-W^{\prime}({\tilde z})$
would be $W^{\prime}_A=2v/(1+\kappa)$,
$W^{\prime}_B=\kappa W^{\prime}_A$. We consider here $v=10$, and
$\kappa=2.$ (For PMMA, $\kappa_A=6.3$ and for PS, $\kappa_B\simeq 2.52$,
so that $\kappa=2.5.)$  Therefore, the magnitude of the
dimensionless electric fields would be
$W^{\prime}_A=20/3$, $W^{\prime}_B=40/3.$
Given a system in which there {\em are} free ions,
we must solve the Poisson-Boltzmann and Laplace
equations, Eqs.~(\ref{pb}) and (\ref{laplace4}).
This is done 
numerically by means of a procedure described in Appendix~II.
The solution yields the potential, electric field, and charge density as
a function of position ${\tilde z}.$

%%%%%%%%%%%%%%%%%%%%%%%%%%%%%%%%%%%%%%%%%%%%%%%%%%
\subsubsection{Local Neutrality}
%%%%%%%%%%%%%%%%%%%%%%%%%%%%%%%%%%%%%%%%%%%%%%%%

Consider the case in which {\em each} A lamella is electrically neutral. 
It is clear that in this case a dipole moment scaling with the distance 
between plates can only be obtained when the lamellae are oriented 
perpendicular to the plates. Thus, the applied field will
 favor this orientation, and we expect the critical field needed
to bring it about will be  reduced by the presence of the
free ions.

We examine a system in which the wavelength of the lamellar structure is 
$\lambda=d/4$ because the effect of the ions is more easily seen in thin 
films. There are four layers of A, each of thickness $d/8$, three layers 
of B, each of thickness $d/8$ and two layers of B next to the two plates, 
each of thickness $d/16$. Were there no ions, the dimensionless potential 
$W$ would be as shown in Fig.~\ref{noions}. The presence of free 
ions does not noticeably affect the potential, even for values 
of the charge density corresponding to $r^2=100.$ Presumably this is 
because the charge separation can never exceed $\lambda/2$ and so is never 
macroscopic. The effect on the electric 
field $\sim W'$, is small, but is discernible for a large charge density 
$r^2=100$, and is shown in Fig.~\ref{elocal}. The non-zero charge density, 
$\sim W''$, can also be discerned, and is shown in Fig.~\ref{rholocal}. 
One sees that the variation of the charge density across each A lamella is 
essentially linear for small charge densities, being positive on one side 
of the lamellae and negative on the other. This is a result of the imposed 
charge neutrality of each lamella. For large charge densities, deviation 
from this linear behavior is expected.

%%%%%%%%%%%%%%%%%%%%%%%%%%%%%%%%%%%%%%%%%%%%%%%%%%
\subsubsection{Global Neutrality}
%%%%%%%%%%%%%%%%%%%%%%%%%%%%%%%%%%%%%%%%%%%%%%%%

Next consider the case in which the A layers are not constrained to be 
locally neutral; the system is only subject to the weaker constraint that 
the system be globally neutral. While the perpendicular orientation of 
lamellae is unaffected by the difference between these constraints, the 
parallel orientation is greatly affected. Each A lamella will now exhibit 
a non-zero charge density profile that is a function of the distance to 
the plates. Thus, there will be a dipole moment that scales with system 
size in the parallel orientation as well as in the perpendicular 
orientation. As a consequence, the latter has no obvious advantage over 
the former, and it is unclear whether there will be any reduction in the 
field needed to bring about a reorientation of lamellae.

Results for the potential, electric field and charge density in the 
parallel orientation are shown in Figs.~\ref{potglobal}, \ref{eglobal}, 
and \ref{rhoglobal}, respectively, for various charge densities.  For the small density of 
charges, $r^2=0.01$, the potential is imperceptibly perturbed from the 
case of no ions, shown in Fig.~\ref{noions}, and the electric field hardly 
differs from the values $20/3$ and $40/3$ which would be obtained were 
there no ions.  We also note that the electric field when $r^2 = 100$ is not symmetric 
about the midpoint of the sample, ${ z}/d=0.5$, because there is no 
symmetry on interchange of positive ions, which are fixed, and negative 
counterions, which are free to move.

As the density of free ions is increased, the charges are free to migrate 
towards the bounding electrodes and to concentrate at the surfaces of the 
lamellae. The charge densities are much larger than in the locally neutral 
case. (Compare the scales of the charge density in Fig.~\ref{rhoglobal} 
with those in the locally neutral case, Fig.~\ref{rholocal}). One notes 
from Fig.~\ref{rhoglobal} that, for large charge densities, $r^2$, the 
mobile negative charges are essentially depleted from the lamellae close 
to the negatively charged plate leaving behind a positive charge density 
with dimensionless value equal to $r^2$. The charge separation is of order of 
the entire film thickness, $d$. As a consequence, the electric field 
becomes increasingly screened, an effect most pronounced deep in the 
interior of the system and which spreads toward the two electrodes with 
increasing charge density. This is not difficult to understand. If one 
groups the layers of opposite charge in nesting pairs, with the outermost 
pair being at ${z/d}=0$ and $1$, then the system resembles a set of nested 
capacitors, of alternating polarity, in series. As each capacitor 
contributes almost no field outside of its plates, the electric field can 
only be large near the two bounding surfaces of the entire system. In the 
limit of very large densities, it is clear that the charge is concentrated 
in the A lamella nearest the plates, and the field and associated
potential drop is only significant in the B layers adjacent to the plates.

%%%%%%%%%%%%%%%%%%%%%%%%%%%%%%%%%%%%%%%%%%
\subsection{The Perpendicular Orientation}
%%%%%%%%%%%%%%%%%%%%%%%%%%%%%%%%%%%%%%%%%%%

For the case in which the lamellae are perpendicular to the plates, we 
must solve the Poisson-Boltzmann equations in two dimensions. Because of 
periodicity and symmetry, it is sufficient to solve them only in an area 
of length $d$ in the $z$ direction and of length $\lambda/2$ in the $x$ 
direction, from the middle of one A lamella to the middle of the adjoining 
B. Equipotential contours are shown in Fig.~\ref{contour} for an applied 
potential $v=10$ and with a charge density $r^2=100$ . The boundary 
between A and B regions runs vertically through the center of the figure 
with the A region to the left, the B to the right. One sees immediately 
that there is little variation of the electric field from A to B regions. 
This is due to the fact that the lamellae are quite narrow compared to 
their height. But one also notices that the equipotentials are very 
different from the equally spaced ones which would characterize the 
potential in the absence of ions. Because the top, positive potential, 
plate attracts the negative mobile counterions, the electric field is 
screened effectively near it and the potential changes slowly; most of the 
change in potential occurs near the bottom plate. 

As the density of free ions is increased, the $A$ lamellae act more and
more like conductors, screening their interiors from the external
electric field, just as in the parallel orientation. But in
the perpendicular orientation, they provide a direct path between
capacitor plates with the consequences that the external voltage source must place an
increasing amount of charge on the plates in order 
to establish the requisite potential difference, and
must expend ever more energy in order to do so.

%%%%%%%%%%%%%%%%%%%%%%%%%%%%%%%%%%%%%%%%%%%
\subsection{The Transition between Orientations}
%%%%%%%%%%%%%%%%%%%%%%%%%%%%%%%%%%%%%%%%%%%

With the solution of the Poisson-Boltzmann equations in hand for the two 
configurations, the corresponding free energies can be evaluated for any 
given external voltage, and in particular for voltages at which the 
perpendicular orientation of lamellae becomes of lower free energy than 
the parallel configuration. We calculate the free energies per unit area 
$f_{\rm el}^\perp$ and $f_{\rm el}^\parallel$ for these two configurations 
and define
\begin{equation}
  \Delta f\equiv  f_{\rm el}^\perp- f_{\rm el}^\parallel.
  \end{equation}
Further we note that in the perpendicular configuration ${\cal S}_A={\cal 
S}_B={\cal S}_{\rm plate}$ because we have assumed that the A and B 
lamellae are of equal thickness. We assume that the surface fields cause 
the plates to prefer the less-polar polymer comprising the B lamellae, as 
in the experiments of Xu et al. \cite{xu04}, so that in the parallel 
configuration ${\cal S}_A=0$ and ${\cal S}_B=2{\cal S}_{\rm plate}$. From 
a comparison of the total free energies of Eq.~(\ref{dimless}) for the two 
orientations, it follows that the perpendicular configuration is favored 
when
\begin{equation}
  \Delta f\leq- \left(\frac{\beta^2\lambda
e^2}{2\epsilon_0\kappa_A}\right)(\gamma_A-\gamma_B).
  \end{equation}
In Fig.~\ref{deltaflocal} a plot of $\Delta f$ vs. the dimensionless 
applied voltage $v$ is presented for the case in which the A lamellae are 
locally neutral. The lamellar periodicity $\lambda$ and film thickness $d$ 
are taken to be $\lambda=d/10,$ which corresponds to the system of Xu et 
al \cite{xu04}. Results for four different values of $r^2$, the 
dimensionless density of the mobile ions, are shown.

The critical value of the external voltage needed to align the lamellae is 
simply determined by the value of $v$ at which a horizontal line drawn at 
the value $\Delta f=- (\beta^2\lambda 
e^2/2\epsilon_0\kappa_A)(\gamma_A-\gamma_B)$ intersects the curve $\Delta 
f(v)$. In particular, consider the experiments of Xu et al. \cite{xu07} 
for which A is PMMA, B is PS, and $\gamma_A-\gamma_B\approx0.5\times 
10^{-3}$J\,m$^{-2}.$ With an absolute temperature $T=430$K, a periodicity 
$\lambda\approx 30\,$nm and $\kappa_A=6.3$, the characteristic surface energy 
$2\epsilon_0\kappa_A/\beta^2\lambda e^2$ is $5.1\times 10^{-6}$J\,m$^{-2}$ 
leading to a dimensionless surface energy difference of $\Delta f=-97$. A 
horizontal line at that value is shown in Fig.~\ref{deltaflocal}. We see that the 
dimensionless voltage needed to bring about alignment is reduced from the 
value without ions of about $v=200$, corresponding to $V_0\simeq 7.5$\,V, 
to about $v=60$, or $V_0\simeq 2.2$\,V, when the density of ions 
corresponds to $r^2=100$. (For a film thickness $d=300\,$nm, the unit of 
number density, $\kappa_A\epsilon_0/\beta e^2 d^2=1.4\times 
10^{20}$\,m$^{-3}$ so $r^2=100$ corresponds to $\rho_+=1.4\times 
10^{22}$\,m$^{-3}$).

This reduction in critical voltage comes from the fact that the 
counterions do little to lower the electrostatic energy of the system when 
the lamellae are parallel to the substrate as well as locally neutral. In 
this case, there is no separation of charge on the order of the film 
thickness. But the ions do a great deal to lower the free energy when the 
lamellae are perpendicular to the plates so that there is a separation of 
charge on the order of the film thickness. Of course, this separation of 
charge decreases the entropy and is therefore opposed by thermal effects. 
In order to determine whether a reorientation will result from the two 
competing effects, one must carry out the calculation explicitly, as we have done, taking 
into account both the decrease of the free energy due to the reduction in 
electrostatic energy and the increase in free energy due to the decrease 
in counterion entropy. As the charge density increases without limit,
the free energy of the perpendicular orientation at any non-zero voltage
decreases without limit so that the curves of free energy difference
approach the ordinate of Fig. 8, and the voltage needed to bring about
reorientation approaches zero.

We next consider the less restrictive case when the system is globally 
neutral. The free energy difference between perpendicular and parallel 
orientations is shown in Fig.~\ref{deltafglobal}. Again there is a 
reduction in the voltage needed to reorient the lamellae,  
from $v \approx 200$ ($r^2=0$, or no ions)  
to $v \approx 160$ for $r^2=100$. This is a much smaller reduction than 
in the locally neutral case for which the reorienting potential was
driven down to
$v \approx 60$ for $r^2=100$. The reason that there is {\em any} reduction in the
reorienting voltage in the globally neutral case is 
subtle for there is a 
macroscopic separation of charge in {\em both} orientations. However, 
while the charge in the perpendicular alignment can be separated by a 
distance $d$, in the parallel alignment it can only be separated by a 
smaller distance $d-\lambda/2$ when the B layer is next to the plates. 
Hence, the ions lower the free energy of the perpendicular configuration 
more than that of the parallel one with a consequent reduction in critical 
voltage. Because the difference between charge separations of $d$ and 
$d-\lambda/2$ is relatively larger the smaller the value of $d$, we expect 
that the relative reduction of the potential needed to reorient the film 
is greater for thinner films. We have verified this conclusion by 
calculating results for films in which $d$ was reduced to 4$\lambda$. 
Whereas in the $d=10\lambda$ film there is a reduction in the 
reorientation potential of about $17$\% when the dimensionless surface 
field is equal to 100 and the density of ions is $r^2=100$, in the thinner 
film the reduction for the same surface field and ion density is about 
$32$\%.

Were the A material preferentially adsorbed to the plates, there would be 
a macroscopic separation of charge of order $d$ in both orientations. In 
fact, the polarization would be somewhat larger in the parallel 
orientation in which the free charges could coat the plates entirely as 
contrasted with the perpendicular configuration in which they could coat 
only half the area of the plates. Thus we would expect that the addition 
of ions actually enhances the parallel orientation with respect to the 
perpendicular and would cause the critical voltage to {\em increase} 
rather than decrease. We have verified that this is indeed the case, 
although the increase in critical voltage is not large. We also expect the 
effective surface tension of the A block with the substrate to be reduced 
by an amount $\epsilon_0\kappa_AV_0^2/\lambda_D$. This expression is 
identical to the reduction of the liquid-solid interfacial tension of 
polar liquids in external potentials (electrowetting) \cite{mugele2007}.

%%%%%%%%%%%%%%%%%%%%%%%%%%%%%%%
\section{Conclusions}
%%%%%%%%%%%%%%%%%%%%%%%%%%%%%%%

We have considered a system composed of lamellae with two different 
dielectric constants and in which one kind, the A lamellae, contain mobile 
negative ions in a smeared continuum background of positive ones. We 
calculated the effect of these ions on the free energy when the lamellae 
are oriented in the direction of the applied field and compared it to that 
when the lamellae are oriented parallel to the two external electrodes, 
which is the orientation favored by surface interactions. We have assumed 
that the system is in thermal equilibrium so that the comparison of these 
free energies determines the magnitude of the applied voltage necessary to 
bring about the reorientation of the lamellae.

The effect of the ions can be quite significant, but we have shown that it 
depends on several factors. The ions have their largest effect if the 
system is locally neutral, {\em i.e.} they neutralize the background 
charge of each of the A lamellae of the film. In that case, a macroscopic 
separation of charge, with its concomitant lowering of energy, can only be 
brought about if the lamellae undergo the desired reorientation. We find 
that for the surface interactions commonly encountered, the voltage needed 
to bring about the reorientation could be reduced by more than half.  
These results are in fine qualitative agreement with the experiments of Kohn et al.
\cite{kohn09} whose system was one of local neutrality. By applying alternating electric fields at 
various frequencies, they could explicity determine the effect of
the free ions as contrasted with those originating from the different dielectric constants.
They indeed found that at lower frequencies at which the effects of the free ions 
become important, the orienting effect of the electric field was enhanced 
\cite{kohn09,tsori2003}.

For systems in which  
the mobile ions can move from one A lamella to another so that it
is only restricted to be globally neutral,  the effect of the ions is predicted
to be smaller than 
in the locally neutral case, and to depend on other factors 
as well. In particular, if the system is globally neutral, we predict that 
the effect is greater if the surface interactions prefer the B (no ions) 
lamellae, for then the separation of charge in the perpendicular 
orientation is larger than that in the parallel one. The relative 
reduction in the reorienting potential should be the greater the thinner 
the film. If the surface interactions prefer the A layers, then the 
presence of ions can actually increase the voltage needed to bring about 
the reorientation.

Insight into the question of whether the lamellae of a particular system
are locally or globally 
neutral can be obtained by altering the plates so that they prefer A in 
one experiment and B in another. A locally neutral system will be 
relatively unaffected by such a change, while a globally neutral one would 
be significantly changed. An alternative is to solubilize a known amount 
of ions in the pure polymers. Integration of the current that passes 
through the polymer at a given external potential allows one to infer the 
fraction of immobile ions and the solubilization energy.

We have noted that separation of charge which occurs on the scale of the 
film thickness decreases the entropy making reorientation more difficult. 
Hence a reduction of temperature certainly makes it easier to bring about 
the reorientation. In our calculation, this is reflected in the definition 
of the dimensionless measure $r$, Eq.~(\ref{rr}), which shows that a 
decrease in temperature is equivalent to an increase in ionic charge 
density.

At constant temperature, all properties of the system depend upon the 
ratio of the ion density to the applied potential, $\rho_+/V_0$, as seen, 
from Eq.~(\ref{r2v}). Hence for a system with a given ion density at a 
fixed temperature, any ``effective'' dielectric constant must depend upon 
the applied voltage itself, and is therefore not a particularly useful 
concept.

Finally the effects of the ions in our system cannot be attributed to a 
change in the strength of the interaction between components, so whatever 
role such an effect may play in diblock copolymer systems \cite{wang08}, 
it is supplementary to the equilibrium effects considered here.

In sum, we have calculated the change in the aligning potential that can 
be expected upon the introduction of ions into a patterned system of 
different dielectric material \cite{tsori03} such as block copolymer, and 
have highlighted some of the issues to be addressed in order to exploit 
such systems for further applications.

\bigskip

%%%%%%%%%%%%%%%%
{\it Acknowledgments~~~}
%%%%%%%%%%%%%%%
We gratefully acknowledge useful correspondence with A. B\"{o}ker, T. 
Russell, S. Stepanow, T. Thurn-Albrecht and J.-Y. Wang. This work was 
supported by the U.S.-Israel Binational Science Foundation under Grant No. 
2006055 (DA, MS), by the National Science Foundation under Grant No. 
DMR-0803956 (MS), the Israel Science Foundation under Grants No. 284/05 
(YT) and No. 231/08 (DA), and by the German Israeli Foundation under Grant 
No. 2144-1636.10/2006 (YT).

\newpage
%%%%%%%%%%%%%%%%%%%%%%%%%%%%%%%%%%%%%%%%%%%%%%%%%%%
\section*{Appendix I: Derivation of the Electrostatic Energy}
\renewcommand{\theequation}{I.\arabic{equation}}
  \setcounter{equation}{0}
%%%%%%%%%%%%%%%%%%%%%%%%%%%%%%%%%%%%%%%%%%%%%%%%%%%%
To obtain the partition function of the system, one must first obtain the
energy of a given configuration of charges.
The electric energy is specified by the number density of the $N_{-}$ mobile negative
charges
\begin{equation}
\label{hatrho}
{\hat\rho}_-({\bf r})=\sum_{i=1}^{N_-}\delta({\bf r}-{\bf r}_i),
\end{equation}
where ${\bf r}_i$ is the position vector of the $i$'th  charge.
The energy is obtained in the standard way and leads to the usual result for a
linear dielectric
  \begin{equation}
  \label{energy}
  U=\frac{1}{2}\epsilon_0\int \kappa({\bf r})[\nabla V({\bf r})]^2d{\bf r}.
  \end{equation}
where the potential, $V[{\hat\rho}_-]$, is a functional of
the number density via the Poisson equation
\begin{equation}
\label{realpoisson}
\nabla^2V({\bf r})=-\frac{1}{4\pi\kappa({\bf
 r})\epsilon_0}{\hat\rho}_-({\bf r}).
\end{equation}
This energy, however, does not include the work that the external electric 
power source, which is part of the system, must do in order to keep 
constant the potential difference between the two electrodes 
\cite{landau84}.  We include this contribution to the energy and denote 
the result for the total energy of the system $U_{tot}.$ Using the fact 
that one plate is at potential zero, and the other at potential $V_0$, the 
decrease in total energy can be written
\begin{eqnarray}
\label{r1}
U_{tot}-U&=&-\int V_0\sigma({\bf r}_{\parallel})d{\bf r}_{\parallel},\\
\label{r2}
 &=&\int V({\bf r}_{\parallel})D_z({\bf r}_{\parallel})d{\bf r}_{\parallel},\\
\label{r3}
 &=&\int \nabla\cdot[V({\bf r})D({\bf r})]d{\bf r},\\
\label{r4}
 &=&\int[\nabla V({\bf r})]\cdot D({\bf r})d{\bf r}+
\int V({\bf r})\nabla\cdot {\bf D}({\bf r})d{\bf r},\\
\label{r5}
 &=&-\int\epsilon_0\kappa({\bf r})[\nabla V({\bf r})]^2d{\bf r}+
\int_A V({\bf r})e[\rho_+-{\hat\rho}_-({\bf r})]d{\bf r}
 \end{eqnarray}
In Eq.~(\ref{r1}) above, the charge density of the upper plate is denoted
$\sigma({\bf r}_{\parallel})$ with ${\bf r}_{\parallel}$ the position
vector in the plane of the plate. The Maxwell equation $\nabla\cdot{\bf
D}=e\rho_{free}$ has been used to express this charge density in terms of
the $z$ component of the displacement field just outside the plate, $D_z$,
which appears in Eq.~(\ref{r2}), and again to relate the divergence of the
displacement field in the bulk dielectric, Eq.~(\ref{r4}), to the free
charge there, Eq.~(\ref{r5}). The second integral in that equation is over the regions of A lamellae which contain the charges. Combining the expressions of
Eqs.~(\ref{energy}) and (\ref{r5}) we obtain for the energy of the system the
expression
\begin{equation}
\label{tildeU}
U_{tot}[{\hat\rho}_-]=-\frac{1}{2}\int\epsilon_0\kappa({\bf r})
[\nabla V({\bf r})]^2d{\bf r}+
\int_A V({\bf r})e[\rho_+-{\hat\rho}_-({\bf r})]d{\bf r}
\end{equation}
which, like the potential $V[{\hat\rho}_-],$ is a functional of the charge
density ${\hat\rho}_-({\bf r}).$
The partition function in the canonical ensemble is
\begin{equation}
Z(T,\Omega,V_0,N_+,{\cal S}_{plate})=\frac{1}{N_-!\lambda_B^{3N_-}}
\int\Pi_{j=1}^{N_-}d{\bf r}_j\,
\exp[-\beta U_{tot}[{\hat\rho}_-]],
\end{equation}
where $\lambda_B$ is the de Broglie wavelength of the negative charges.

There are many methods to obtain from the exact partition function the
mean-field approximation to the free energy, such as that
pioneered by Edwards \cite{edwards65} and used commonly
since\cite{borukhov00,fredrickson06, negele88}. They all lead to the result
\begin{eqnarray}
F_{\rm el}(T,\Omega,V_0,N_+,{\cal S}_{\rm plate})
&=&U_{tot}[\rho_-({\bf r})]-TS,\\
\label{end}
&=&-\frac{1}{2}\epsilon_0\int\kappa({\bf r})[\nabla V({\bf
r})]^2d{\bf r}+e\int_A V({\bf r})[\rho_+-\rho_-({\bf r})]\,d{\bf r} \nonumber \\
                    &+&k_BT\int_A\,
\rho_-({\bf r})\ln[\rho_-({\bf r})/\rho_+]\,d{\bf r},
\end{eqnarray}
where $\rho_-({\bf r})$, in contrast to ${\hat\rho}_-({\bf r})$ of
  Eq.~(\ref{hatrho}), is an ensemble-average density
\begin{equation}
\rho_-({\bf r})=\rho_+\Omega\frac{\exp[\beta eV({\bf r})]}
{\int_A d{\bf r}\exp[\beta e V({\bf r})]} \qquad {\bf r}\ {\rm in\  A}
\end{equation}
and is the source of the potential via the Poisson equation
\begin{equation}
\nabla^2V({\bf r})=-\frac{1}{4\pi\kappa({\bf
 r})\epsilon_0}\rho_-({\bf r}).
\end{equation}
The mean-field free energy, Eq.~(\ref{end}), is seen to be simply the 
total electrostatic energy, Eq.~(\ref{tildeU}), but as a function of this 
average charge density, supplemented by the contribution to the entropy of 
the mobile counterions.

\newpage
%%%%%%%%%%%%%%%%%%%%%%%%%%%%%%%%%%%%%%%%%%%%%%%%%%%%%%%
\section*{Appendix II: Numerical Solutions}
\renewcommand{\theequation}{II.\arabic{equation}}
  \setcounter{equation}{0}
%%%%%%%%%%%%%%%%%%%%%%%%%%%%%%%%%%%%%%%%%%%%%%%%%%%%%%%

The rescaled electrostatic free energy $f_{\mathrm{el}}$ defined in 
Eq.~(\ref{scaledfel}) can be written in terms of the rescaled potential 
$W$ by using the relation $\rho_-({\bf r}) = \rho_{+}\theta \exp(W({\bf 
r}))$ to eliminate the explicit dependence on $\rho_{-}({\bf r})$ in 
Eq.~(\ref{fmft}).  The result, after changing to rescaled variables and 
neglecting a constant which contributes equally to the parallel and 
perpendicular configurations, is
\begin{eqnarray} f_{\mathrm{el,global}} = 
&-&\frac{1}{2}\int_{\tilde{x}=0}^{\lambda/2d} \int_{\tilde{z}=0}^{1} 
\frac{\kappa(\tilde {x},\tilde{z})}{\kappa_{A}} \left[ {\bf \nabla} 
W(\tilde{x},\tilde{z}) \right]^{2} d\tilde{x} d\tilde{z} \nonumber \\ &+& 
r^{2} \int \int_{A} W(\tilde{x},\tilde{z}) d\tilde{x} d\tilde{z} \nonumber 
\\ &+& \frac{1}{2} r^{2} \left(\frac{\lambda}{2d}\right) \ln \theta
\end{eqnarray}
This expression is valid for both the globally neutral parallel 
configuration and the perpendicular configuration.  In the case of locally 
neutral parallel lamellae, each lamella labeled by $i$ has its own 
constant $\theta_{i}$ and the rescaled free energy is

\begin{eqnarray} f_{\mathrm{el,local}} = 
&-&\frac{1}{2}\int_{\tilde{x}=0}^{\lambda/2d} \int_{\tilde{z}=0}^{1} 
\frac{\kappa(\tilde{x},\tilde{z})}{\kappa_{A}} \left[ {\bf \nabla} 
W(\tilde{x},\tilde{z}) \right]^{2} d\tilde{x} d\tilde{z} \nonumber \\ &+& 
r^{2} \int \int_{A} W(\tilde{x},\tilde{z}) d\tilde{x} d\tilde{z} \nonumber 
\\ &+&r^{2} \left(\frac{\lambda}{2d}\right) 
\sum_{i}(\frac{\Omega_{A,i}}{\Omega})\ln \theta_{i},
\end{eqnarray}
where $\Omega_{A,i}$ is the volume taken up by the $i$th lamella within 
the integration region $\Omega$.  These rescaled free energies can be 
calculated once the rescaled potential $W$ and the constant $\theta$ (or 
constants $\theta_{i}$ in the case of locally neutral lamellae) have been 
determined by numerically solving discretized versions of the equations
\begin{align} 
&{\tilde\nabla}^2W({\tilde x},{\tilde z})
~~=~~-r^2\left(1-\theta_{i}{\rm e}^{W({\tilde x},{\tilde z})}\right) 
~~~~~~~~~&{\rm in\ lamella\ A}_{i}& \nonumber \\
&{\tilde\nabla}^2W({\tilde x},{\tilde z})~~=~~~~~~~~~0 ~~~~~~~&{\mbox{in B 
regions}}\
\end{align}
subject to appropriate boundary conditions.  In the case of parallel 
lamellae where $W$ is a function of $\tilde{z}$ only, the following 
boundary conditions hold in both the locally and globally neutral cases:
\begin{eqnarray}
\label{lamellarbc1}
W(0) &=& 0 \\
\label{lamellarbc2}
W(1) &=& v \\
\label{lamellarbc3}
\left.\kappa^{-} W'(\tilde{z}_{i})\right|_{\rm left} &=& 
\left.\kappa^{+}W'(\tilde{z}_{i})\right|_{\rm right} \\
\label{lamellarbc4}
\left.W(\tilde{z}_{i})\right|_{\rm left} &=& 
\left.W(\tilde{z}_{i})\right|_{\rm right},
\end{eqnarray}
where the AB interfaces are indexed by $i$ and $\kappa^{-}$ and 
$\kappa^{+}$ are the relative dielectric constants on the two sides of the 
interface at $\tilde{z}_{i}$.  In the case of globally neutral lamellae, 
the unknown $\theta$ is determined by the boundary condition
\begin{eqnarray}
W'(0) = W'(1)
\end{eqnarray}
In the case of locally neutral lamellae, each unknown $\theta_{i}$ is 
determined by a boundary condition describing the neutrality of the $i$th 
A-lamella:
\begin{eqnarray}
W'(a_{i}) = W'(b_{i}), \qquad 1 \le i \le n,
\end{eqnarray}
where $n$ is the number of A-lamellae and $a_{i}$ and $b_{i}$ are the 
scaled positions of the edges of the $i$th such lamella. For the perpendicular 
configuration, the boundary conditions are
\begin{eqnarray}
W(\tilde{x},0) = 0 \\
W(\tilde{x},1) = v \\
\left.\kappa_{A}\frac{\partial W}{\partial \tilde{x}}\right|_{\tilde{x} = 
\lambda/4d^{-}} = \kappa_{B}
\left.\frac{\partial W}{\partial \tilde{x}}\right|_{\tilde{x} = \lambda/4d^{+}},
\qquad 0 \le {\tilde z} \le 1 \\
\label{symmetrybc1}
\left.\frac{\partial W}{\partial \tilde{x}}\right|_{\tilde{x} = 0} = 0,
\qquad 0 \le {\tilde z} \le 1 \\
\label{symmetrybc2}
\left.\frac{\partial W}{\partial \tilde{x}}\right|_{\tilde{x} = \lambda/2d} = 0,
\qquad 0 \le {\tilde z} \le 1 \\
\left.W(\lambda/4d, \tilde{z})\right|_{\rm left} = 
\left.W(\lambda/4d, \tilde{z})\right|_{\rm right},  \qquad 0 \le {\tilde z} \le 1 \\
\kappa_{A} \int_{0}^{\lambda/4d} 
\left(\left.\frac{\partial W}{\partial \tilde{z}}\right|_{\tilde{z}=1}-
\left.\frac{\partial W}{\partial 
\tilde{z}}\right|_{\tilde{z}=0}\right)\,d\tilde{x} \nonumber \\
\label{perpendicularneutralitybc}
+ \kappa_{B} \int_{\lambda/4d}^{\lambda/2d} 
\left(\left.\frac{\partial W}{\partial 
\tilde{z}}\right|_{\tilde{z}=1}-\left.\frac{\partial W}{\partial \tilde{z}}
\right|_{\tilde{z}=0} \right)\, d\tilde{x} = 0,
\end{eqnarray}
where $\tilde{x} = 0$ is in the middle of an A lamella, so that $\tilde{x} 
= \lambda/4d$ is the location of the AB interface.  
Eqs.~(\ref{symmetrybc1}) and (\ref{symmetrybc2}) enforce the symmetry of 
the system with respect to reflection about the center of each A or B 
lamella, respectively.  Eq.~(\ref{perpendicularneutralitybc}) determines 
the value of $\theta$ by enforcing the neutrality of the system.

\newpage
%%%%%%%%%%%%%%%%%%%%%%%%%%%%%%%%%%%%%%%
\bibliographystyle{aipnum4-1}

\newpage
%%%%%%%%%%%%%%%%%%%%%%%%%%%%%%%%
\section*{Figure Captions}
%%%%%%%%%%%%%%%%%%%%%%%%%%%%%%%%%

\begin{itemize}

\item[]{\bf Figure 1.~~}
Parallel case, no ions ($r^2=0$). The dimensionless applied potential
is $v=10,$ and $\kappa=\kappa_A/\kappa_B=2.$ The sample
thickness and wavelength are related by
$d=4\lambda.$ The dimensionless potential, $W\equiv \beta e V$ is
plotted vs. the dimensionless coordinate $ z/d$.
A-type regions are shown in white, B regions in gray.

\item[]{\bf Figure 2.~~}
Parallel case, local neutrality. The dimensionless applied potential
is $v=10,$ and $\kappa=\kappa_A/\kappa_B=2.$ The sample
thickness and wavelength are related by
$d=4\lambda.$ The dimensionless electric field, $-W^{\prime}$, is
plotted vs. the dimensionless coordinate $ z/d$.
A-type regions are shown in white, B regions in gray. The case for no ions, $r^2=0$ is shown with solid lines, while $r^2=100$
is shown with a dotted line.

\item[]{\bf Figure 3.~~}
Parallel case, local neutrality. The dimensionless total charge density, $\ -W^{\prime\prime}$,
is plotted for $v=10$, $\kappa=2.$ The sample
thickness and wavelength are related by
$d=4\lambda.$  Solid line, $r^2=50$; dashed line,
 $r^2=100$.

\item[]{\bf Figure 4.~~}
Parallel case, global neutrality.
The dimensionless potential $W$ for $v = 10$, $\kappa =
2$, and with $r^2$ varied.  Solid line, $r^2 = 0.01$; dashed line,
$r^2 = 50$;
dotted line, $r^2 = 100$. The sample
thickness and wavelength are related by
$d=4\lambda$.

\item[]{\bf Figure 5.~~}
Parallel case, global neutrality. The dimensionless
electric field, $-W^{\prime},$ is plotted for  $v = 10$, $\kappa =
2$, and with $r^2$ varied.  Solid line, $r^2 = 0.01$; dashed line,
$r^2 = 100$. The sample
thickness and wavelength are related by
$d=4\lambda.$

\item[]{\bf Figure 6.~~}
Parallel case, global neutrality. The dimensionless charge density,
$-W^{\prime\prime}$, is plotted
for $v = 10$, $\kappa = 2$, and
with $r^2$ varied.  Solid line, $r^2 = 50$;
dotted line, $r^2 = 100$. The sample
thickness and wavelength are related by
$d=4\lambda$.

\item[]{\bf Figure 7.~~}
Perpendicular case. Contours of constant
dimensionless potential, $W$, at intervals of 1.0
are  plotted in the ${ x/d},{z/d}$ plane for $v=10$ and $r^{2}=100$. As $\lambda=d/4$, we
need only plot ${x/d}$ from 0.0 to $\lambda/2d=0.125.$ The
interface between A and B lamellae is at ${x/d}=0.0625$ with
the A region to the left and B to the right of it. Areas between equipotentials are filled
with different shades of gray for clarity.

\item[]{\bf Figure 8.~~}
The difference between the dimensionless electrostatic free energies per unit
area (without surface fields) of the perpendicular orientation and the locally
neutral parallel orientation is shown as a function of dimensionless applied voltage
$v$ for four values of ion number density; $r^2=0.01$ (solid line),
$50$ (dashed line), $100$ (dotted line), and $1000$ (dotted-dashed line).
The sample
thickness and wavelength are related by
$d=10\lambda.$ The horizontal line is at the value $\Delta f=-97$ (see text).

\item[]{\bf Figure 9.~~}
The difference between the dimensionless electrostatic free energies
per unit
area (without surface fields) of the perpendicular orientation and the globally
neutral parallel orientation is shown as a function of dimensionless applied voltage
$v$ for four values of ion number density; $r^2=0.01$, (solid line)
$50$ (dashed line), $100$ (dotted line), and $1000$ (dotted-dashed line).
The sample
thickness and wavelength are related by
$d=10\lambda.$ The horizontal line is at the value $\Delta f=-97$ (see text).
\end{itemize}

\newpage

%%%%%%%%%%%%%%%%%%%%%%%%%%%%%%%%%%%%%%
\begin{center}
%{\bf Figures}
\end{center}
\renewcommand{\thefigure}{\arabic{figure}}
  \setcounter{figure}{0}
%%%%%%%%%%%%%%%%%%%%%%%%%%%%%%%%%%%%%%%%%%

%Figure1
%noionspotential
\begin{figure}[htp]
  \begin{center}
   %{\resizebox{3.5in}{!}{\includegraphics{fig1.eps}}}
     {\resizebox{3.5in}{!}{\includegraphics{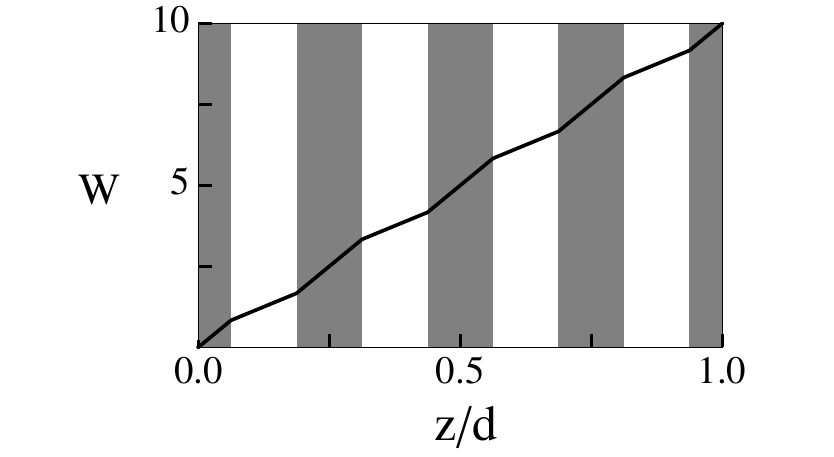}}}
    \caption{\label{noions}}
  \end{center}
\end{figure}
\vskip 5truecm
%Figure2
%noionspotential
\begin{figure}[htp]
  \begin{center}
    % {\resizebox{3.5in}{!}{\includegraphics{fig2.eps}}}
     {\resizebox{3.5in}{!}{\includegraphics{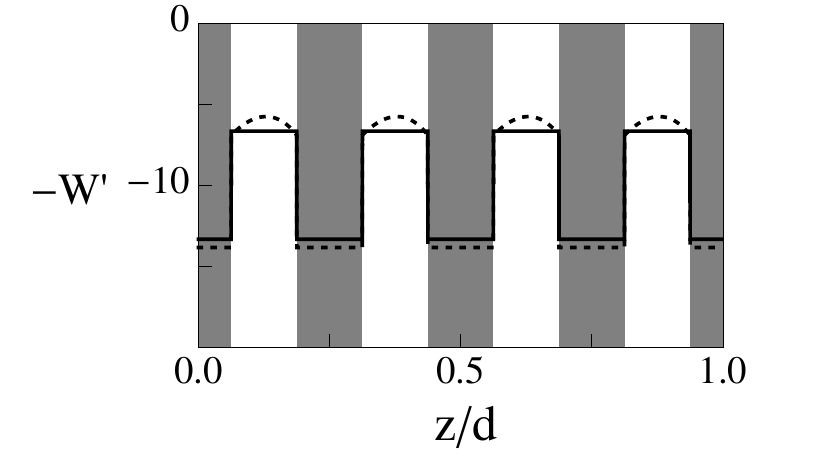}}}
    \caption{\label{elocal}}
  \end{center}
\end{figure}

\vskip 5truecm
%Figure3
%chargedensityneutral
\begin{figure}[htp]
  \begin{center}
    % {\resizebox{3.5in}{!}{\includegraphics{fig3.eps}}}
     {\resizebox{3.5in}{!}{\includegraphics{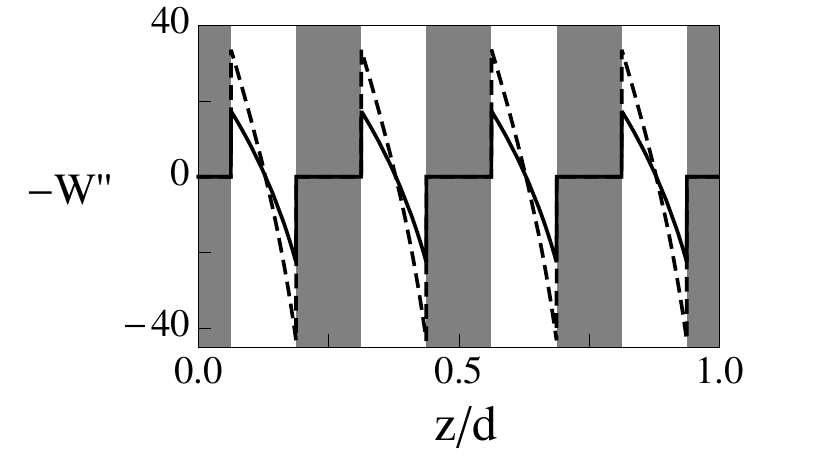}}}
    \caption{\label{rholocal}}
  \end{center}
\end{figure}
\clearpage

%Figure4
%potentialglobal
\begin{figure}[htp]
  \begin{center}
 % {\resizebox{3.5in}{!}{\includegraphics{fig4.eps}}}
    {\resizebox{3.5in}{!}{\includegraphics{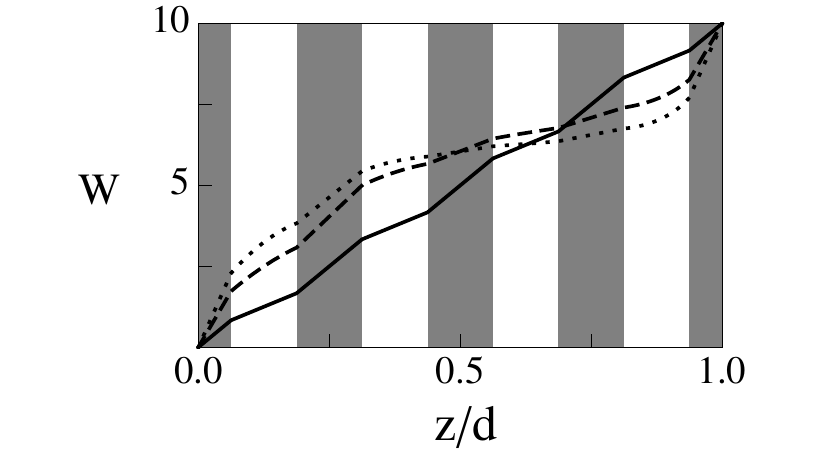}}}
    \caption{\label{potglobal}}
  \end{center}
\end{figure}
%\hfill

\vskip 5truecm

%Figure5
%efieldglobal
\begin{figure}[htp]
     \begin{center}
%{\resizebox{3.5in}{!}{\includegraphics{fig5.eps}}}
{\resizebox{3.5in}{!}{\includegraphics{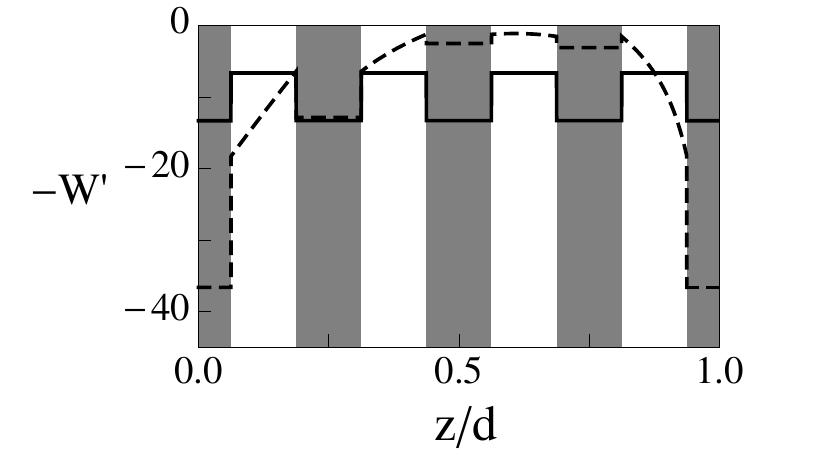}}}
\caption{
 \label{eglobal}}
     \end{center}
 \end{figure}
\clearpage

%Figure6
%chargedensityglobal
\begin{figure}[htp]
    \begin{center}
%{\resizebox{3.5in}{!}{\includegraphics{fig6.eps}}}
{\resizebox{3.5in}{!}{\includegraphics{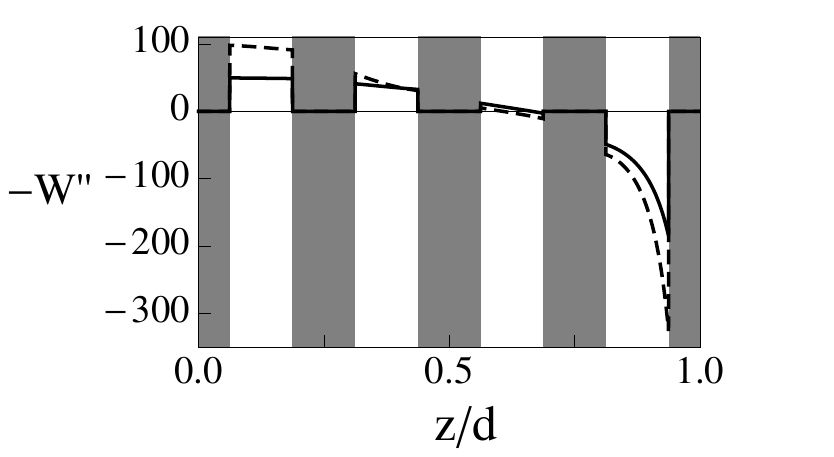}}}
\caption{
\label{rhoglobal}}
     \end{center}
\end{figure}

\vskip 5truecm

%Figure7
%contourplot
\begin{figure}[htp]
    \begin{center}
%{\resizebox{2.5in}{!}{\includegraphics{fig7.eps}}}
{\resizebox{3.5in}{!}{\includegraphics{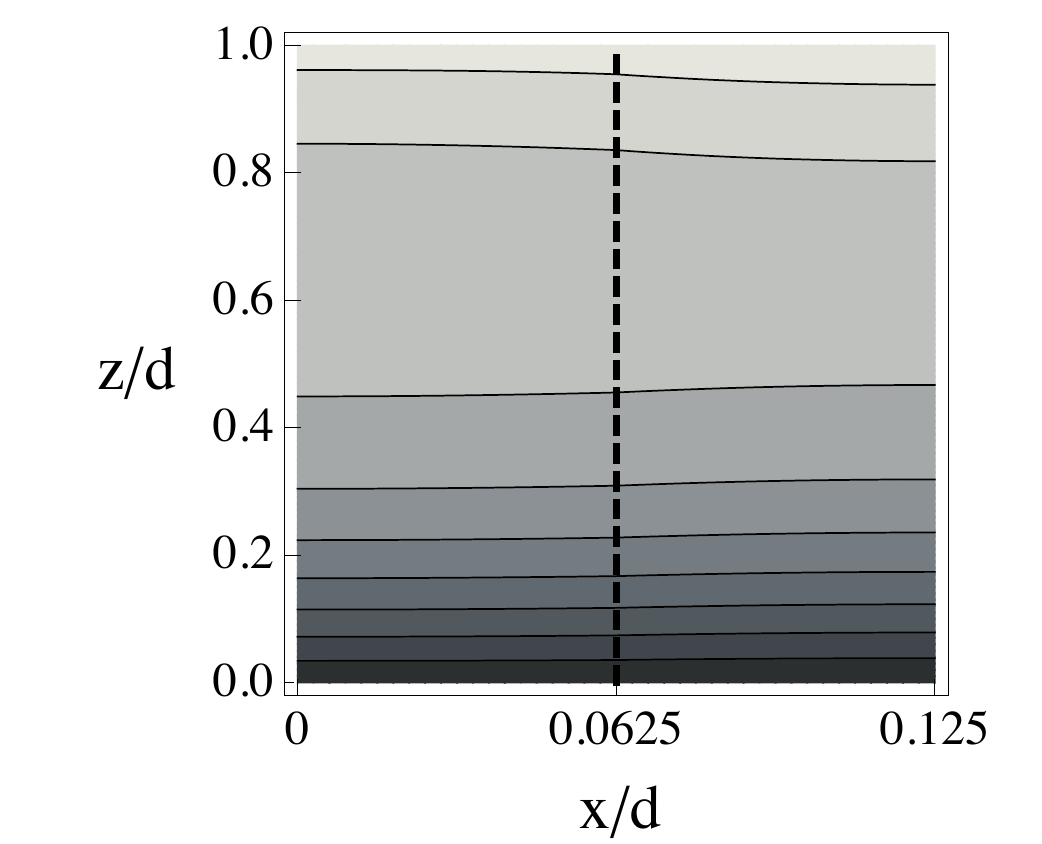}}}
\caption{
\label{contour}}
     \end{center}
\end{figure}
\clearpage

%Figure8
%deltaflocal
\begin{figure}[htp]
  \begin{center}
    %{\resizebox{3.5in}{!}{\includegraphics{fig8.eps}}}
   {\resizebox{3.5in}{!}{\includegraphics{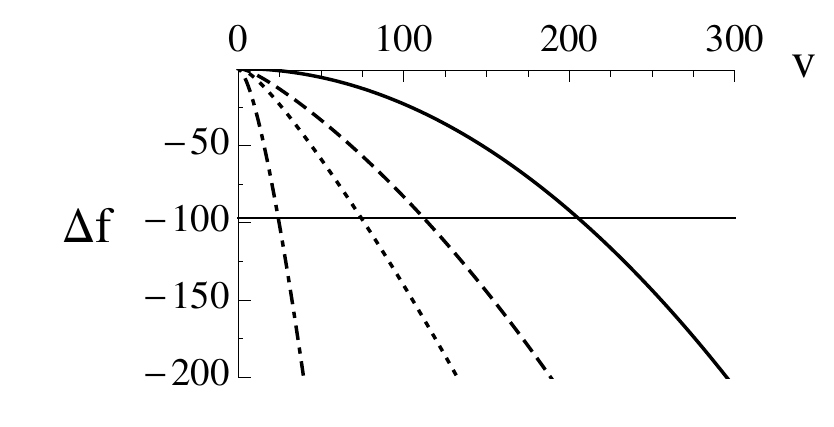}}}
    \caption{\label{deltaflocal}}
  \end{center}
\end{figure}

\vskip 5truecm

%Figure9
%deltafglobal
\begin{figure}[htp]
  \begin{center}
    % {\resizebox{3.5in}{!}{\includegraphics{fig9.eps}}}
    {\resizebox{3.5in}{!}{\includegraphics{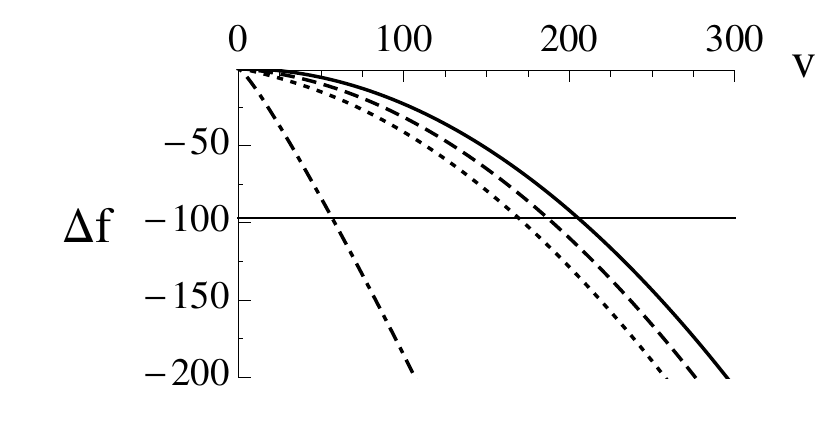}}}
    \caption{\label{deltafglobal}}
  \end{center}
\end{figure}
\clearpage

%\bibliography{polymer-09}
\end{document}